\documentclass{article}
\usepackage{graphicx}
\usepackage{url}
\usepackage{amsfonts}
\usepackage{amssymb}
\usepackage{latexsym}

\def\F{{\sc f}}
\def\T{{\sc t}}
\newenvironment{pf}{\unskip{\bf Proof:}}{\unskip{\hfill $\Box$}}

\newcommand{\lemlab}[1]{\label{lemma:#1}}

\newcommand{\tablab}[1]{\label{tab:#1}}
\newcommand{\figlab}[1]{\label{fig:#1}}
\newcommand{\seclab}[1]{\label{section:#1}}

\newcommand{\lemref}[1]{\ref{lemma:#1}}

\newcommand{\tabref}[1]{\ref{tab:#1}}
\newcommand{\figref}[1]{\ref{fig:#1}}
\newcommand{\eqref}[1]{(\ref{eq:#1})}
\newcommand{\secref}[1]{\ref{section:#1}}

\newtheorem{theorem}{Theorem}
\newtheorem{lemma}{Lemma}

{\catcode`\@=11
\gdef\setft#1#2#3{%
\def\@oddfoot{
{\setbox0=\hbox{#1}
\setbox1=\hbox{#3}
\ifdim\wd0>\wd1
\dimen0=\wd0
\box0\hfil#2\hfil\hbox to\dimen0{\hfil\hfil\box1}
\else \dimen0=\wd1
\hbox to\dimen0{\box0\hfil }\hfil#2\hfil\box1 \fi
}}} }

\def\complaint#1{}
\def\withcomplaints{
\newcounter{mycomplaints}
\def\complaint##1{\refstepcounter{mycomplaints}%
\ifhmode%
\unskip%
{\dimen1=\baselineskip \divide\dimen1 by 2 %
\raise\dimen1\llap{\tiny -\themycomplaints-}}\fi%
\marginpar{\tiny [\themycomplaints]: ##1}}%
}

\let\oldendpf=\endpf
\def\endpf{\oldendpf\par\medskip}

\title{\bf PushPush is NP-hard in 2D}
\author{%
Erik D. Demaine \and Martin L. Demaine\thanks{
Dept.\ Comput\ Sci., Univ.\ Waterloo,
Waterloo, Ontario N2L 3G1, Canada.
\texttt{\{eddemaine, mldemaine\}@\penalty \exhyphenpenalty uwaterloo.ca}.
}
\and
Joseph~O'Rourke\thanks{
Dept.\ Comput.\ Sci., Smith Col\-lege, North\-ampton,
MA 01063, USA.
\texttt{orourke@\penalty \exhyphenpenalty cs.smith.edu}.
Supported by NSF grant CCR-9731804.}
}

\begin{document}
\maketitle
\begin{abstract}
We prove that a particular pushing-blocks puzzle is intractable in 2D,
improving an earlier result that established intractability in
3D~\cite{ppnph-os-99}.
The puzzle, inspired by the game {\em PushPush},
consists of unit square blocks on an integer lattice.
An agent may push blocks (but never pull them) in attempting
to move between given start and goal positions.
In the PushPush version, the agent can only push one block at
a time, and moreover, each block, when pushed, slides the
maximal extent of its free range.
We prove this version is NP-hard in 2D by reduction from SAT.
\end{abstract}

\section{Introduction}
There are a variety of ``sliding blocks'' puzzles whose
time complexity has been analyzed.
One class, typified by the 15-puzzle so heavily
studied in AI, permits an outside
agent to move the blocks.
Another class falls more under the guise of
motion planning.
Here a robot or internal agent plans a
path in the presence of movable obstacles.
This line was initiated by a paper of Wilfong~\cite{w-mppmo-91},
who proved NP-hardness of a particular version in which the
robot could pull as well as push the obstacles, which were
not restricted to be squares.
Subsequent work sharpened the class of problems by weakening
the robot to only push, never pull obstacles, and by 
restricting all obstacles to be unit squares.
Even this version is NP-hard when some blocks may be
fixed to the board (made unpushable)~\cite{do-mpams-92}.

One theme in this research has been to establish stronger degrees
of intractability, in particular,   
to distinguish between NP-hardness and PSPACE-completeness, 
the latter being the stronger claim.  
The NP-hardness proved in~\cite{do-mpams-92} was 
strengthened to PSPACE\--completeness in an unfinished 
man\-u\-script \cite{bos-mpams-94}.  
More firm are the results on
Sokoban, a
computer game that restricts the pushing robot to only push one block at
a time, and requires the storing of (some or all) blocks into 
designated ``storage
locations.''
This game was proved NP-hard in~\cite{dz-sompp-95},
and PSPACE-complete by Culberson~\cite{c-spc-99}.  

Here we emphasize another theme: finding a nontrivial version of
the game that is {\em not\/} intractable.  To date only the most uninteresting
versions are known to be solvable in polynomial time, for example,
where the robot's path must be monotonic~\cite{do-mpams-92}.
We explore a different version,
again inspired
by a computer game, PushPush.\footnote{
        The earliest reference we can find to the game
        is a version written for the Macintosh by
        Alan Rogers and C.M. Mead III, Copyright 1994,
        \url{http://www.kidsdomain.com/down/mac/pushpush.html}.
        Another version for the Amiga was written by
        Luigi Recanatese in 1997,
        \url{http://de.aminet.net/aminet/dirs/game_think.html}.
}  
The key difference is that
when a block is pushed, it necessarily slides the full extent
of the available empty space in the direction in which it was
shoved. This further weakens the robot's control, and the
resulting puzzle has certain polynomial characteristics.
It was established in~\cite{ppnph-os-99} that the problem is
intractable in 3D, but its status in 2D was left open in that paper.
Here we settle the issue by extending the reduction to 2D.

\section{Problem Classification}
\seclab{Classification}
The variety of pushing-block puzzles may be classified by
several characteristics:

\begin{enumerate}
\item Can the robot pull as well as push?
\item Are all blocks unit squares, or may they have different shapes?
\item Are all blocks movable, or are some fixed to the board?
\item Can the robot push more than one block at a time?
\item Is the goal for the robot to move from $s$ to $t$,
or is the goal for the robot to push blocks into storage locations?
\item Do blocks move the minimal amount, exactly how far they
are pushed, or do they slide the maximal amount of their
free range?
\item The dimension of the puzzle: 2D or 3D?
\end{enumerate}

If our goal is to find the weakest robot and most
unconstrained puzzle conditions that still lead to intractability, 
it is reasonable to consider robots who can only push~(1),
and to restrict all blocks to be unit squares~(2), 
as in \cite{do-mpams-92,dz-sompp-95,c-spc-99}, for
permitting robots to pull, and permitting blocks of other shapes,
makes it relatively easy to construct intractable puzzles.
It also makes sense to explore the goal of simply finding a path~(5)
as in \cite{w-mppmo-91,do-mpams-92}, rather than
the more challenging task of
storing the blocks as in Sokoban~\cite{dz-sompp-95,c-spc-99}.

Restricting attention to these choices
still leaves a variety of possible problem definitions.
If the robot can only move one block at a time, then the
distinction between all blocks movable and some fixed essentially disappears,
because $2 \times 2$ clusters of blocks are effectively fixed to a robot
who can only push one.
If all blocks are movable and the robot can push more than one
at a time, then the blocks should be confined to a
rectangular frame. 

The version explored in this paper superficially seems that it might
lend itself to a polynomial-time algorithm: the robot can only push
one block~(4), all blocks are pushable~(3), and finally, the
robot's control over the pushing is further weakened by condition~(6):
once pushed, a block slides (as without friction) the maximal
extent of its free range in that direction.
Allowing the robot to move in 3D gives it more ``power'' than
it has in 2D, so the natural question after~\cite{ppnph-os-99}
is to explore the weaker option of condition~(7):  PushPush
in 2D.

Because our proof is a direct extension of the proof for 3D
in~\cite{ppnph-os-99}, we repeat through Section~\secref{SAT.Reduction} the 3D
construction in that paper
(with a minor simplification),
and in Section~\secref{Crossovers} show a similar
reduction holds in 2D as well.  (Both reductions are from SAT, i.e.,
satisfiability of formulas in conjunctive normal form.)  A summary of related
results is presented in the final section.

\section{Elementary Gadgets}
First we observe, as mentioned above, that any $2 \times 2$ cluster of 
movable blocks is forever frozen to a PushPush robot, for there
is no way to chip away at this unit.  This makes it easy to
construct ``corridors'' surrounded by fixed regions to guide
the robot's activities.  We will only use corridors of
width~1 unit, with orthogonal junctions of degree two, three,
or four.  We can then view a particular PushPush puzzle as
an orthogonal graph, whose edges represent the corridors, understood
to be surrounded by sufficiently many $2 \times 2$ clusters to render any
movement outside the graph impossible.
We will represent movable blocks in the corridors or at corridor
junctions as circles.

We start with three elementary gadgets.

\subsection{One-Way Gadget}

A {\em One-Way\/} gadget is shown in
Fig.~\figref{One-Way}a.
It has these obvious properties:
\begin{figure}[htbp]
\centering
\includegraphics[width=0.7\textwidth]{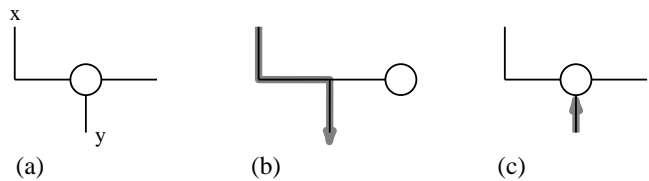}
\caption{One-Way gadget: permits passage from $x$ to $y$ but
not from $y$ to $x$.}
\figlab{One-Way}
\end{figure}

\begin{lemma}
In a One-Way gadget,
the robot may travel from point $x$ to point $y$,
but not from $y$ to $x$.
(After travelling from $x$ to $y$, however,
the robot may subsequently return from $y$ to $x$.)
\lemlab{One-Way}
\end{lemma}
\begin{pf}
The block at the degree-three junction may be pushed into
the storage corridor when approaching from $x$,
as illustrated in Fig.~\figref{One-Way}b,
but the block may not be budged when approaching from $y$
(Fig.~\figref{One-Way}c).
\end{pf}

\subsection{Fork Gadget}
The {\em Fork gadget\/}\footnote{
        This is a simplification of the functionally
        equivalent gadget used in~\cite{ppnph-os-99}.
}
shown in
Fig.~\figref{Fork}a presents the robot with a binary choice,
the proverbial fork in the road:
\begin{figure}[htbp]
\centering
\includegraphics[width=\textwidth]{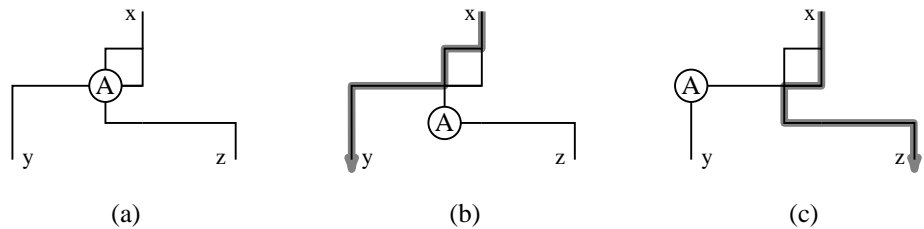}
\caption{Fork gadget: Robot may pass (b) from $x$ to $y$
or (c) from $x$ to $z$, but each seals off the other possibility.}
\figlab{Fork}
\end{figure}

\begin{lemma}
In a Fork gadget,
the robot may travel from point $x$ to $y$, or from
$x$ to $z$, but if it chooses the former it cannot later
move from $y$ to $z$, and if it chooses the latter it cannot
later move from $z$ to $y$.
(In either case, the robot may reverse its original path.)
\lemlab{Fork}
\end{lemma}
\begin{pf}
Fig.~\figref{Fork}b shows the only way for the robot to pass
from $x$ to $y$.  Now the corridor to $z$ is permanently sealed
off by block $A$.
Fig.~\figref{Fork}c shows the only way to move from $x$ to $z$,
which similarly seals off the path to $y$.
\end{pf}

Note that in both these gadgets, the robot may reverse its
path, a point to which we will return in Section~\secref{Crossovers}.

\subsection{3D Crossover Gadget}
Crossovers are trivial in 3D, as shown in 
Fig.~\figref{3D.crossover}.
\begin{figure}[htbp]
\centering
\includegraphics{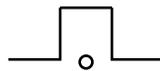}
\caption{3D crossover. The central small circle is a wire
orthogonal to the plane of the figure.}
\figlab{3D.crossover}
\end{figure}

\section{Variable-Setting Component}
The robot first travels through a series of variable-setting
components, each of which follows the structure shown in
Fig.~\figref{Variable}: a Fork gadget, followed by two paths,
labeled {\sc t} and {\sc f}, each
with attached {\em wires} exiting to the right,
followed by a re-merging of the
{\sc t} and {\sc f} paths via One-Way gadgets.
3D crossovers are illustrated in this and subsequent figures
by broken-wire underpasses.
\begin{figure}[htbp]
\centering
\includegraphics[height=6cm]{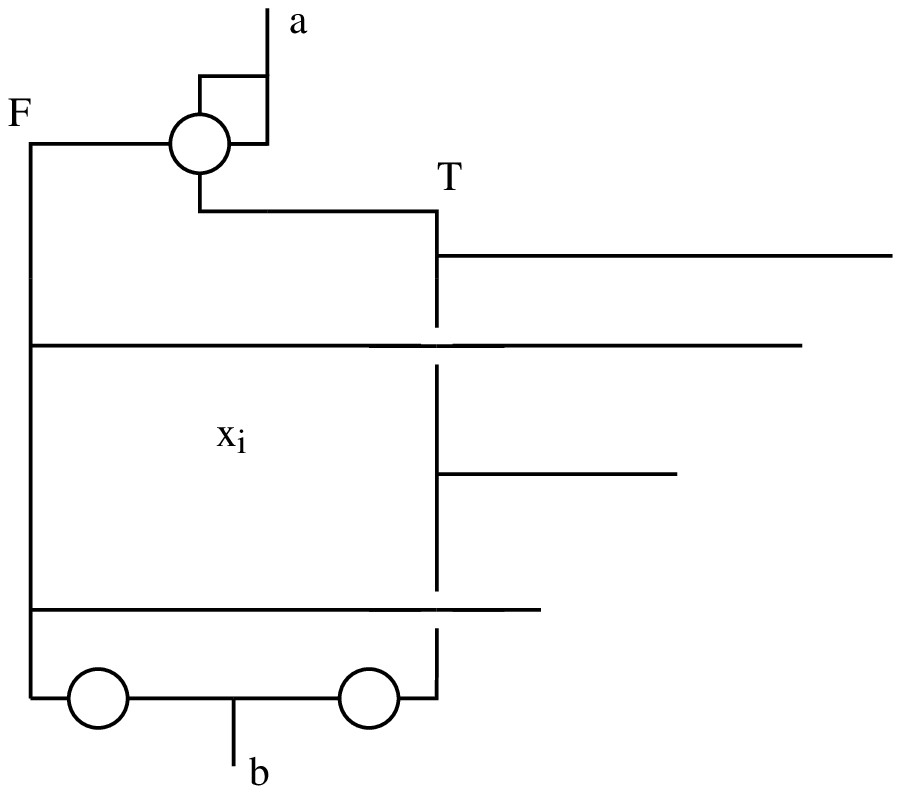}
\caption{(a) Variable $x_i$ component.}
\figlab{Variable}
\end{figure}
\begin{lemma}
The robot may travel from $a$ to $b$ only by choosing either the
{\sc t}-path, or the {\sc f}-path, but not both.
Whichever {\sc t/f}-path is chosen allows the robot to travel down any wires
attached to that path, but down none of the wires attached to the other
path.
\lemlab{Variable}
\end{lemma}
\begin{pf}
The claims follow directly from
Lemma~\lemref{Fork}
and
Lemma~\lemref{One-Way}.
\end{pf}

\section{Clause Component}
The clause component 
shown in Fig.~\figref{Clause}a
cannot be traversed unless one or more blocks (``keys'')
are pushed in
from the left along the attached horizontal wires.
\begin{figure}[htbp]
\centering
\includegraphics[width=0.8\textwidth]{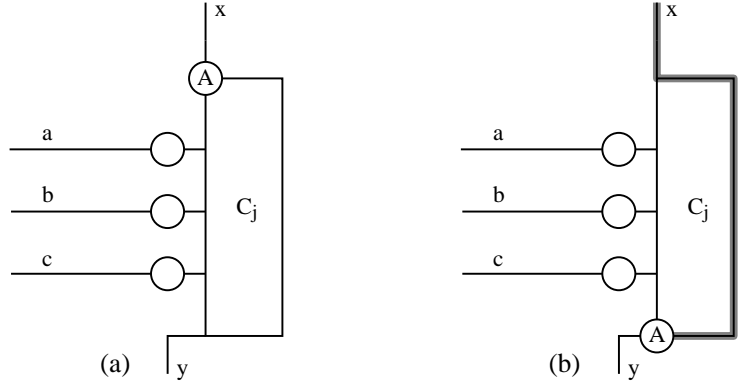}
\caption{(a) Clause $C_j$ component.
(b) Passage thwarted without an ``inserted'' key block.}
\figlab{Clause}
\end{figure}
\begin{lemma}
The robot may only pass from $x$ to $y$ of a clause component
if at least one block is pushed into it along an attached wire
($a$, $b$, or $c$ in Fig.~\figref{Clause}a).
\lemlab{Clause}
\end{lemma}
\begin{pf}
Block $A$ is necessarily pushed by the robot starting at $x$.
This block will clog exit at $y$ (Fig.~\figref{Clause}b)
unless its sliding is stopped
by a block pushed in on an attached wire.
\end{pf}

\noindent
The basic mechanism that gives a clause component its functionality
will be reused in several guises in Section~\secref{Crossovers}, 
so we pause to
redescribe it.  Essentially the component is a {\em lock\/} with three
{\em keys}, which we identify with the three blocks on the
$a$, $b$, and $c$ wires.  A necessarily-pushed block $A$ is
characteristic of locks, as is an alternate path around the
spot(s) where the key(s) come to rest.

\section{Complete SAT Reduction}
\seclab{SAT.Reduction}
The complete construction for four clauses
$C_1 \wedge C_2  \wedge C_3 \wedge C_4$
is shown in 
Fig.~\figref{3D.SAT}.
Two versions of the clauses are shown in the figure:
an unsatisfiable formula (the dark lines),
and a satisfiable formula (including the shaded $x_2$ wire):

\begin{eqnarray}
&(x_1 \vee x_2) \wedge  (x_1 \vee \sim x_2) \wedge  (\sim x_1 \vee x_3) \wedge (\sim x_1 \vee \sim x_3) \\
&(x_1 \vee x_2) \wedge  (x_1 \vee \sim x_2) \wedge  (\sim x_1 \vee x_2 \vee x_3) \wedge (\sim x_1 \vee \sim x_3)
\end{eqnarray}
Here we are using $\sim x$ to represent the negation of the variable $x$.

\begin{figure}[htbp]
\centering
\includegraphics[width=\textwidth]{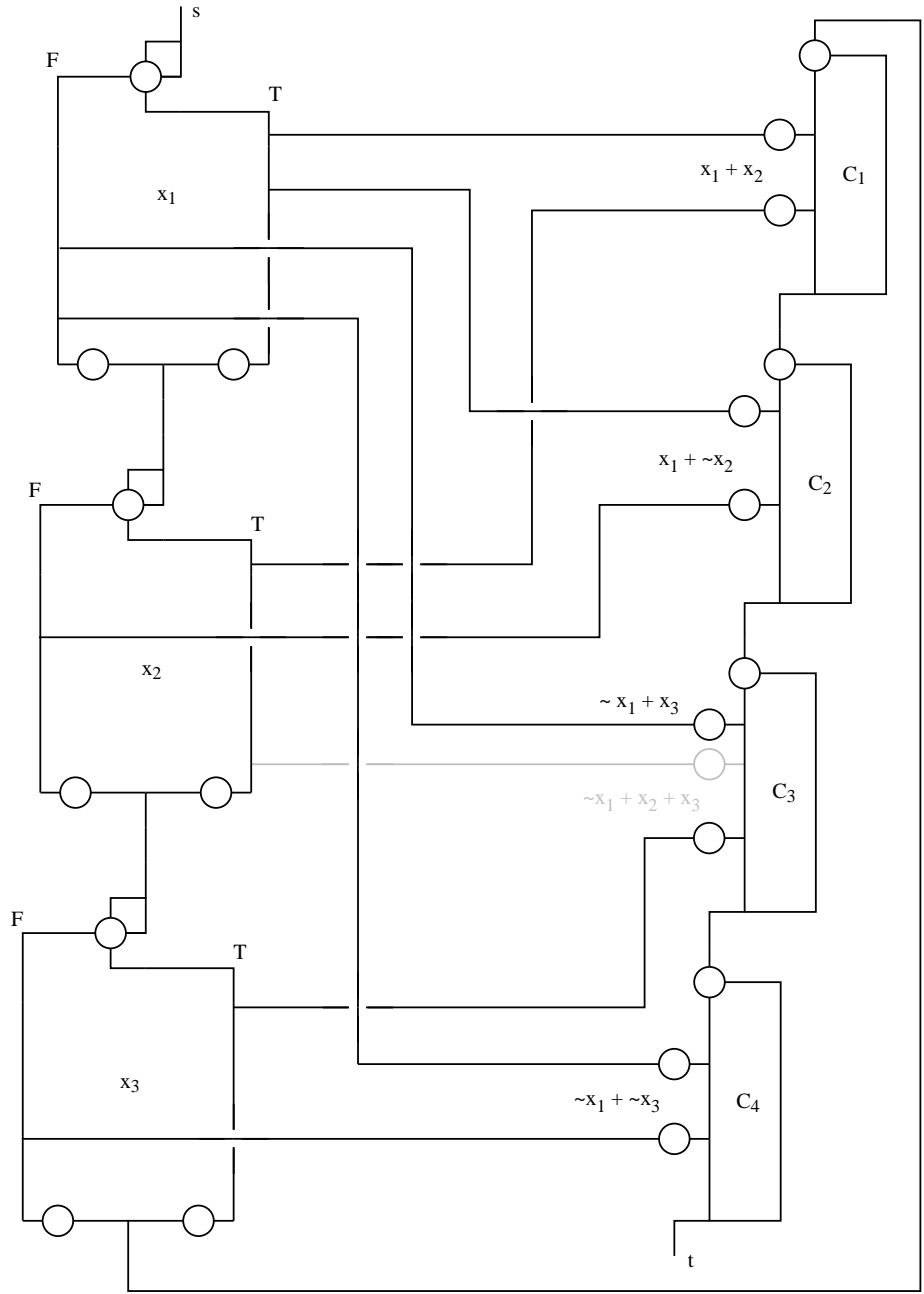}
\caption{Complete construction for the formulas in Eq.~(1) and Eq.~(2)
(including the shaded portion).}
\figlab{3D.SAT}
\end{figure}
A path from $s$ to $t$ in the satisfiable version is illustrated in
Fig.~\figref{3D.SAT.path}.
\begin{figure}[htbp]
\centering
\includegraphics[width=\textwidth]{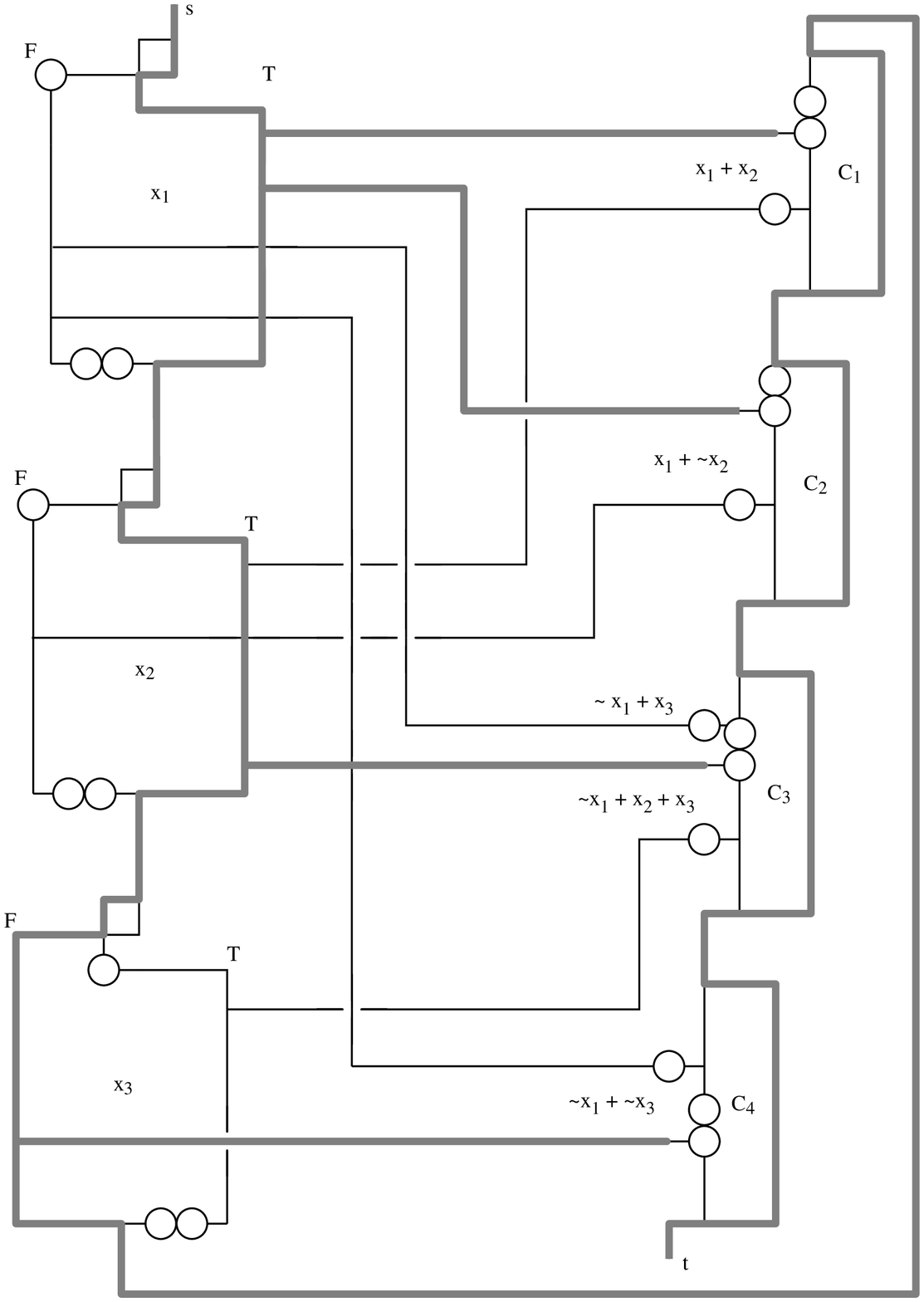}
\caption{Solution path for Eq.~(2).}
\figlab{3D.SAT.path}
\end{figure}

\begin{theorem}
PushPush is NP-hard in 3D.
\end{theorem}
\begin{pf}
The construction clearly ensures,
via Lemmas~\lemref{Variable} and~\lemref{Clause},
that if the simulated Boolean
expression is satisfiable, there is a path from $s$ to $t$,
as illustrated in Fig.~\figref{3D.SAT.path}.
For the other direction, suppose the expression is unsatisfiable.
Then the robot can reach $t$ only by somehow ``shortcutting'' the
design.  The design of the variable components ensures that
only one of the {\sc t/f} paths may be accessed.
The crossovers ensure there is no ``leakage'' between wires.
The only possible thwarting of the design would occur if the
robot could travel from a clause component back to set a variable
to the opposite Boolean value.  But each variable-clause
wire contains a block that prevents any such leakage.
\end{pf}

\section{2D Crossovers}
\seclab{Crossovers}
We now modify the 3D SAT reduction to a 2D SAT reduction
by replacing the 3D crossovers with appropriate 2D crossovers;
it is only here that we deviate substantively from~\cite{ppnph-os-99}.
Note that there are two distinct types of crossovers used in
the construction:
\begin{enumerate}
\item {\em FT-crossovers\/}: A horizontal wire from
an \F-wire in a variable unit crosses the vertical \T-wire of
the same variable unit.
\item {\em VC-crossovers\/}: A horizontal wire from some
variable unit to a clause unit crosses a vertical wire
from some other variable unit to some clause unit.
\end{enumerate}
The FT-crossovers are significantly different from the VC-crossovers
in that the former are traversed in one direction or the other but
never both.  This is because once the robot chooses the \T-wire,
it can never get into the \F-wire, and vice versa (Lemma~\lemref{Variable}).
Thus a limited crossover suffices here.

There is another approach to handling the FT-crossovers:
reduction from
``Planar 3-SAT''~\cite{l-pftu-82}
(as used, e.g., in Dor and Zwick's  NP-hardness proof~\cite{dz-sompp-95})
permits eliminating this type of crossover entirely.
However, we do not pursue that tack, for use of Planar 3-SAT
introduces additional crossovers in the final clause-threading path.
Instead we develop a limited crossover gadget for FT-crossovers, 
which may serve as an introduction to the more
demanding VC-crossovers in Section~\secref{locking.door}.

\subsection{XOR Crossover}

We call the limited crossover gadget an {\em XOR Crossover\/};
it is shown in Fig.~\figref{limited.crossover}(a).
\begin{figure}[htbp]
\centering
\includegraphics[height=0.7\textheight]{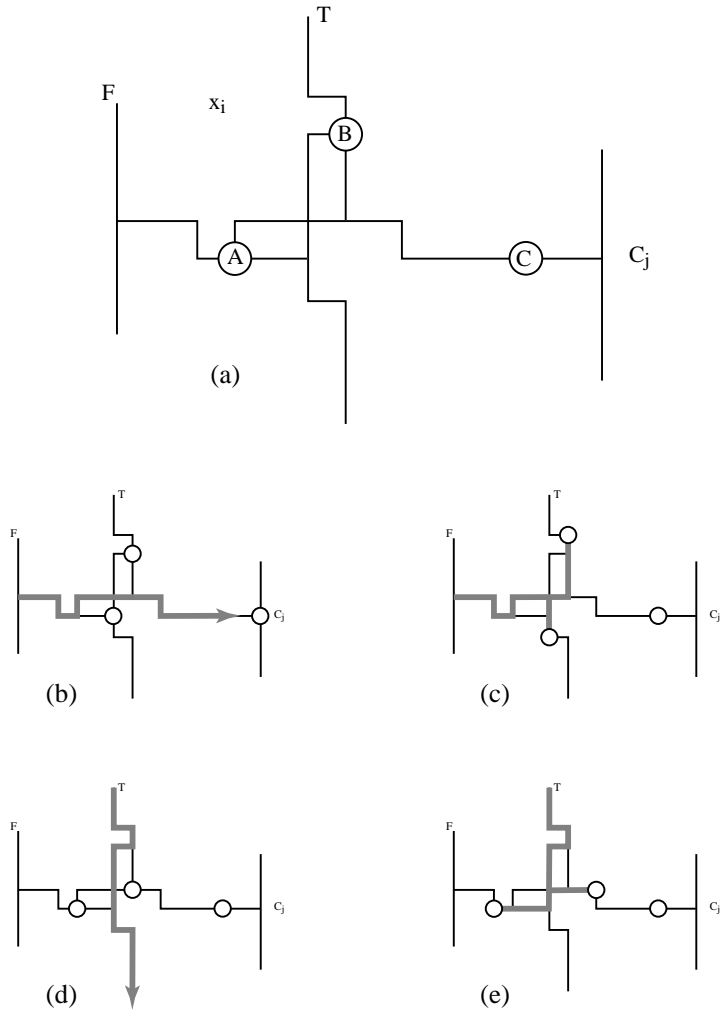}
\caption{(a) XOR Crossover gadget.
(b,c): Horizontal passage;
(d,e): Vertical passage.}
\figlab{limited.crossover}
\end{figure}

\begin{lemma}
An XOR Crossover gadget
permits either horizontal, rightward passage without leakage into the
vertical channel, or vertical, downward passage without leakage into the
horizontal channel, but not both.
In either case, the passage may afterwards be traversed
an arbitrary number of times in the same direction.
\lemlab{limited.crossover}
\end{lemma}
\begin{pf}
Entering the unit from the \F-wire to the left pushes block $A$
into the vertical channel (Fig.~\figref{limited.crossover}(b)),
which, together with block $B$, disallows entry into the vertical
$T$-wire ((c) of the figure).
Similarly, entering the unit from the \T-wire at the top
pushes block $B$ to clog the horizontal channel (d), preventing
leakage into either the \F-wire or the clause unit (e).
\end{pf}

\subsection{Locking Door Unit}
\seclab{locking.door}
VC-crossovers do not have the exclusive-or property that makes
FT-crossovers so easy to handle.
They may need to be traversed in either direction.
However, note that VC-crossovers need only be traversed once
in each of the four directions: from a variable component
$x_i$, to deposit a key into a clause component $C_j$,
and returning back to $x_i$; and later from $x_k$, $k \neq i$, to
some $C_l$, and back to $x_k$.
The design of such a crossover is the most complex part of our
construction, and will proceed in several stages.

The most important gadget used to build this crossover is
one that permits passage in one direction, but then prevents
return in the other direction.  We call this
a {\em Locking Door Unit\/}; it is shown in 
Fig.~\figref{locking.door}(a).
Like a One-Way gadget, it may only be traversed in one direction.
But unlike that gadget, once traversed it becomes a permanently locked
door with respect to both directions.  A more accurate name
for the gadget
might be ``unidirectional, single-use, self-closing and self-locking door.''
\begin{figure}[htbp]
\centering
\includegraphics[height=0.85\textheight]{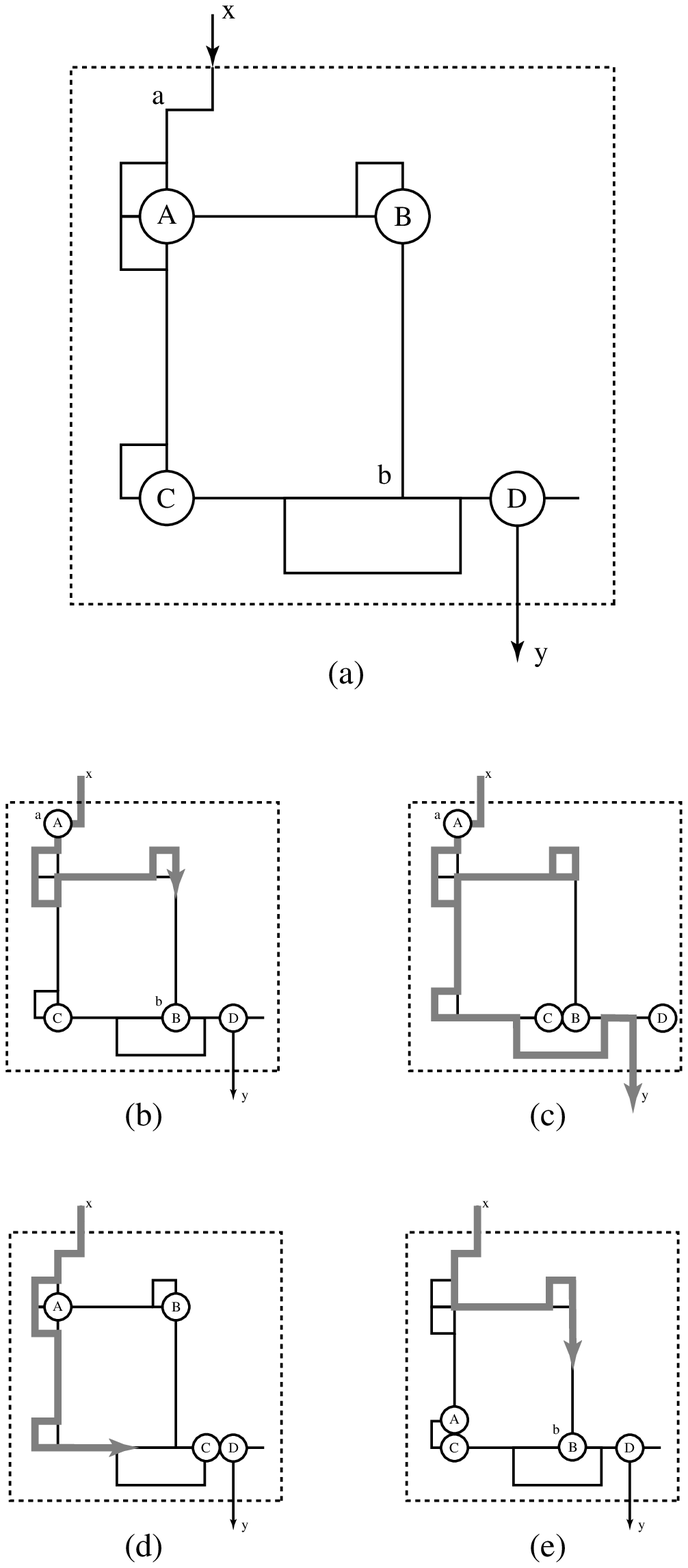}
\caption{(a) Locking door unit.
(b,c): Passage through unit;
(d,e): Bad attempts.}
\figlab{locking.door}
\end{figure}
Note that the unit contains a lock mechanism
(similar to that used in a clause component) centered around
point $b$, which requires the key block $B$ to permit passage.

\begin{lemma}
Upon first encounter,
the robot may pass from $x$ to $y$ through a locking door unit,
and not $y$ to $x$;
but once through, the unit becomes impassable in
either direction.
\lemlab{locking.door}
\end{lemma}
\begin{pf}
We first argue that the unit may be passed through as claimed.
The robot first moves from $x$ to below block $A$, which
it pushes upward to point $a$.  This seals off return passage.
It then pushes block $B$ down to point $b$, as shown in
Fig.~\figref{locking.door}(b).  $B$ serves as a key to an
exit door.  It then ``unlocks'' that door by pushing block
$C$ to abut against $B$.  It may then travel around $CB$,
push block $D$ to the right, and reach the exit $y$.
See (c) of the figure.  Note that it now impossible for
the robot to return from $y$ to $x$; moreover
it is impossible for the robot to retraverse the unit from
$x$ to $y$ again, in both cases because of block $A$ at point $a$.

Next we argue that the above is the only way to traverse the unit.
It may not be traversed from $y$ to $x$ because of the One-Way
gadget represented by block $D$.
We consider the four options of what to do with block $A$ after entering
at $x$:
\begin{enumerate}
\item Push $A$ upward.  This leads to the solution just described.
\item Do not move $A$.  Then $B$ is unreachable, and the only option
is to push block $C$ rightwards.  But without the key $B$ at $b$,
$C$ slides over to abut $D$ and clog the exit.
See~(d) of the figure.
\item Push $A$ downward. Then $A$ slides down to touch $C$.
Now the robot can only push the key $B$ down to $b$ ((e) of the figure),
but then the door cannot be ``unlocked'' with $C$ as that is now
inaccessible.
\item Push $A$ rightward.  Then $A$ hits $B$, and prevents the 
use of the key at $b$.  Now, similar to~(d) of the figure,
$C$ slides all the way to $D$ and prevents exit.
\end{enumerate}
This exhausts the options and establishes the claims of the lemma.
\end{pf}

\noindent
We will use the symbols illustrated in Fig.~\figref{symbols}(a-b)
to represent a locking door before (a) and after (b) passage.
\begin{figure}[htbp]
\centering
\includegraphics[width=5cm]{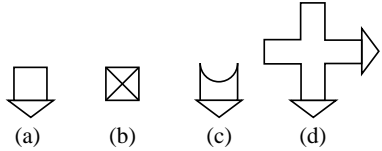}
\caption{Symbols: (a) Locking door prior to passage.
(b) After passage.
(c) Double lock unit prior to passage.
(d) Unidirectional crossover unit.}
\figlab{symbols}
\end{figure}

\subsection{Double Lock Unit}
We next detail a gadget that behaves like a locking door, in that
passage is permitted once in one direction, but which requires
an external key to operate.  We call this a {\em Double Lock Unit}.
It is illustrated in
Fig.~\figref{double.lock}(a).
\begin{figure}[htbp]
\centering
\includegraphics[height=0.85\textheight]{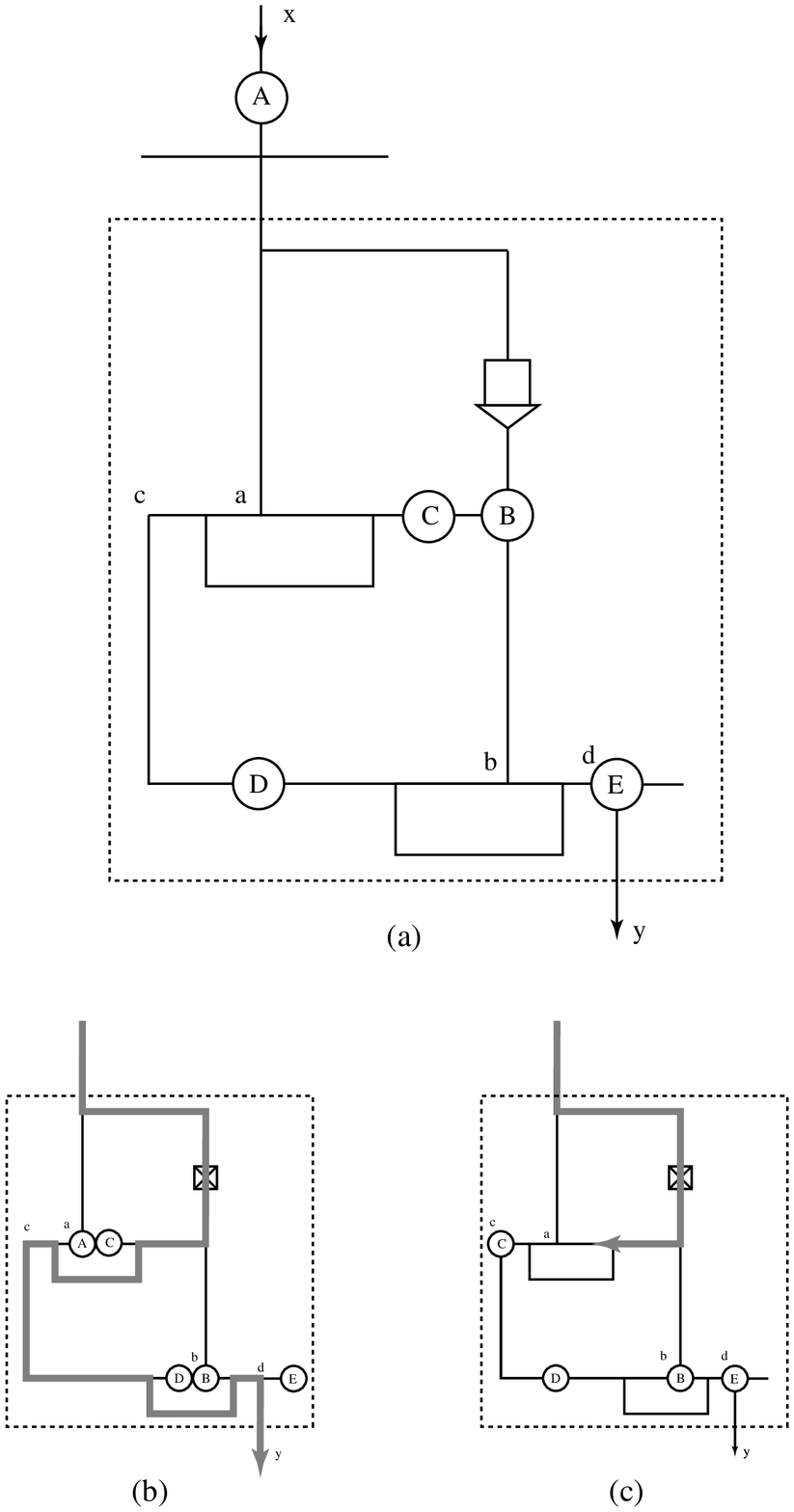}
\caption{(a) Double lock unit. (b) Passage through
with $A$-key. (c) Attempt without $A$-key.}
\figlab{double.lock}
\end{figure}

\begin{lemma}
The robot may pass from $x$ to $y$ through a double lock unit
only if a block $A$ (the external key) is first pushed down
the entrance $x$-wire.
A double lock may not be traversed from
$y$ to $x$,
and once traversed forwards from $x$ to $y$,
the unit becomes impassable in
either direction.
\lemlab{double.lock}
\end{lemma}
\begin{pf}
We call the two locks in the unit the $A$- and $B$-locks.
The former consists of the key block $A$, block $C$,
and the wires near point $a$;
the latter consists of the key block $B$, block $D$,
and the wires near point $b$.
Passage through the unit when there is an external block $A$
is shown in Fig.~\figref{double.lock}(b).
Key block $A$ is pushed to point $a$, and then the robot 
goes through the locking door unit to reach $B$.
This $B$-key is pushed to point $b$.  Now both locks have keys.
The robot then pushes block $C$ leftwards and passes through
the $A$-lock,  pushes block $D$ rightwards and passes
through the $B$-lock, pushes block $E$ out of the way,
and finally exits at $y$.

Suppose there is no external $A$ block, and the robot attempts
to traverse the unit.  To reach $y$, the robot must pass through
the $B$-lock.  If it approaches from the left without first
pushing down the $B$-key, then block $D$ will be pushed rightwards
and clog exit at point $d$.  So the key block $B$ must be pushed
down to point $b$.  The only way to reach $B$ is via the locking
door above it (because block $C$ prevents access from the left).
But passing through the locking door leaves the robot no
alternative but to push $C$ leftwards, as in in Fig.~\figref{double.lock}(c).
This blocks access to the $B$-lock.

The initial position of block $E$
ensures that the unit cannot be traversed from $y$ to $x$ upon
first encounter.  That it cannot be traversed in either direction
after forward passage is clear by inspection of Fig.~\figref{double.lock}(b).
\end{pf}

We will use the symbol shown in Fig.~\figref{symbols}(c) to represent 
a double lock unit prior to passage,\footnote{
        The circular ``bite'' taken out of the
        arrow is to remind us that passage requires an external key.}
and again the `X' in (b) to represent the unit after passage.

\subsection{Unidirectional Crossover Gadget}
We are finally prepared to design a gadget that removes the
exclusive-or limitation of the XOR crossover gadget.
We call this a {\em Unidirectional Crossover gadget\/}; see
Fig.~\figref{uni.cross}(a).
\begin{figure}[htbp]
\centering
\includegraphics[width=0.8\textwidth]{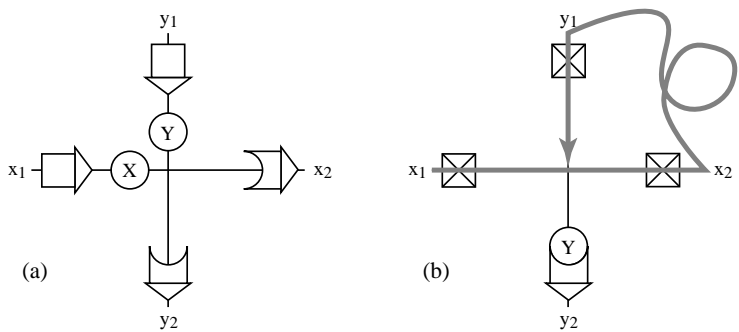}
\caption{(a) Unidirectional Crossover.
(b) No leakage.}
\figlab{uni.cross}
\end{figure}
Call the directed horizontal path $(x_1,x_2)$ the 
{\em forward $x$-path},
and the directed vertical path $(y_1,y_2)$ the {\em forward $y$-path}.
The {\em backward\/} paths are the reverse of these.
The point where the paths meet we call the {\em junction} of the gadget.
The forward $x$-path consists of a locking door
followed by a key block $X$ prior to the junction
and a double lock after the junction; the $y$-path is structured
similarly.

We denote the unidirectional crossover gadget by the symbol in
Fig.~\figref{symbols}(d).

\begin{lemma}
A unidirectional crossover gadget may be traversed 
along the forward $x$-path (but not the backward $x$-path),
and along the forward $y$-path (but not the backward $y$-path),
in either order.
Once either path is traversed, that path (but not the other)
becomes impassable in either direction.
If the junction is approached from $x_1$, then the robot may not
from there reach either $y_1$ or $y_2$ without first passing through
the $x$-path to $x_2$; and symmetrical claims hold after approaching
the junction from $y_1$.
\lemlab{uni.cross}
\end{lemma}
\begin{pf}
The design is symmetrical, so we need only establish its properties
for the $x$-path.
If the robot is at $x_1$, it can only enter the unit by
passing through the locking door, which then is
permanently sealed behind it.
The robot must then
push the $X$ block into the
double lock unit on the $x$-path.  $X$ serves as the external
key for that unit, and permits passage as in Lemma~\lemref{double.lock}.
Thus passage along the forward $x$-path is possible.
The reverse passage is not possible, neither initially
(because a double lock cannot be traversed backwards), nor
after forward passage (because the double lock then becomes closed).

Unlike the XOR crossover, passage along the $x$-path does not
affect the ability to later traverse the forward $y$-path.
Finally, note that when the robot reaches
the junction starting from $x_1$, it may not ``leak''
into the $y$-path: 
its upward movement is stopped by the locking door, and its
downward movement is prevented by the inability to pass through
the double lock without the $Y$-key---the robot is on the ``wrong side''
of the key.
Similarly, if the robot first traverses the $x$-path, and then
reaches the junction via the $y$-path,
then it may not ``leak'' into the $x$-path,
as Fig.~\figref{uni.cross}(b) illustrates.
\end{pf}

\subsection{Bidirectional Crossover Gadget}
We can finally construct a VC-crossover, which we call
a {\em Bidirectional Crossover Gadget\/}; see
Fig.~\figref{bi.cross}.
\begin{figure}[htbp]
\centering
\includegraphics[width=8cm]{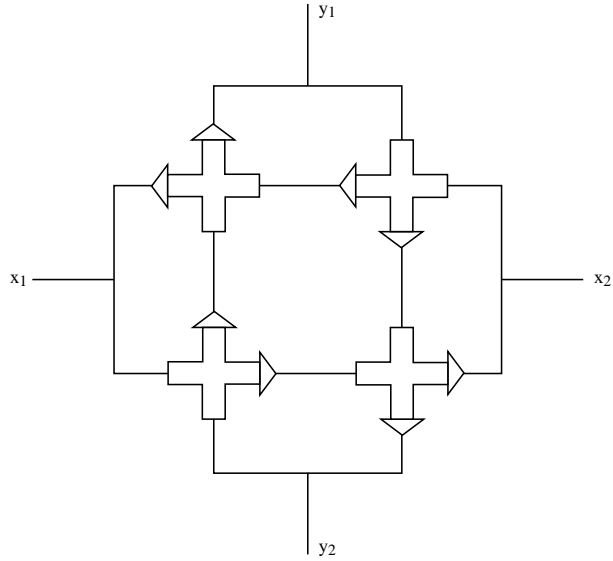}
\caption{Bidirectional crossover.}
\figlab{bi.cross}
\end{figure}
Again call the directed horizontal path $(x_1,x_2)$ the 
{\em forward $x$-path}, and similarly for 
the {\em forward $y$-path} and the {\em backward\/} paths.

\begin{lemma}
A bidirectional crossover may be traversed as many as four
times with any combination, in any order, of the
forward $x$-path, the backwards $x$-path,
the forward $y$-path, and the backwards $y$-path.
Furthermore, no leakage between an x-path and y-path is possible.
\lemlab{bi.cross}
\end{lemma}
\begin{pf}
The claim follows immediately from the construction and
the properties of the unidirectional crossover gadget established
in Lemma~\lemref{uni.cross}.
For example, the forward $x$-path is traversed by passing through
the bottom two unidirectional gadgets.
Later the reverse $y$-path can be traversed by passing through
the left two unidirectional gadgets.  This still leaves it
possible to later traverse the forward $y$-path and the reverse $x$-path.
\end{pf}

\subsection{2D Construction}
Replacing each FT-crossover by an XOR crossover gadget
(Fig.~\figref{limited.crossover}),
and each VC-crossover by a bidirectional crossover gadget
(Fig.~\figref{bi.cross})
completes the 2D construction.
(Note we could replace the FT-crossovers by bidirectional crossover units,
although the extra complexity of this gadget is not needed here.)
The entire construction
may be traversed from $s$ to $t$ iff the represented Boolean
formula is satisfiable.
We have proved:

\begin{theorem}
PushPush is NP-hard in 2D.
\end{theorem}
We leave it an open question whether this theorem can be
strengthened in either direction:
either by proving PushPush is in NP,
in which case it is NP-complete,
or by showning that PushPush is PSPACE-complete.

\section{Summary}
We conclude by summarizing 
in Table~\tabref{results}
previous work according to
the classification scheme offered in Section~\secref{Classification}.
The first four lines show previous results.
The next two are the results from~\cite{ppnph-os-99}.
(The 2D storage result is, incidentally, not difficult.)
The boldface line of the table 
is the result of this paper.

The last two lines list
open problems.
The penultimate line reposes the open question from~\cite{do-mpams-92}:
Is the problem where all blocks are movable and
the robot can push $k$ blocks, sliding the minimal amount,
intractable in 2D?
The last line presents a new open problem,
which we call {\em Push-1\/}:
Is the problem where all blocks are movable, the robot can only
push $1$ block at time, pushing it the minimal amount,
intractable in 2D?
This differs from the problem addressed in this paper only in
altering the sliding from {\em max\/} to {\em min\/}.
Both of these open problems remain candidates for being solvable in polynomial
time, the former because it seems difficult to construct gadgets when
the robot can destroy them, the latter because without sliding the
maximal extent, it seems difficult to design gadgets with the
requisite functionality.

\begin{table}[htbp]
\begin{center}
\begin{tabular}{| c | c | c | c | c | c | c | c |}
        \hline
1 & 2 & 3 & 4 & 5 & 6 & 7 & \mbox{} \\
{\em Push?}
        & {\em Blocks}
        & {\em Fixed?}
        & {\em \#}
        & {\em Path?}
        & {\em Sliding}
        & {\em Dim}
        & {\em Complexity}
        \\ \hline \hline
pull
        & L
        & fixed
        & $k$
        & path
        & min
        & 2D
        & NP-hard \cite{w-mppmo-91}
        \\ \hline
push
        & unit
        & fixed
        & $k$
        & path
        & min
        & 2D
        & NP-hard \cite{do-mpams-92}
        \\ \hline
push
        & unit
        & movable
        & $1$
        & storage
        & min
        & 2D
        & NP-hard \cite{dz-sompp-95}
        \\ \hline
push
        & unit
        & movable
        & $1$
        & storage
        & min
        & 2D
        & PSPACE \cite{c-spc-99}
        \\ \hline \hline
push
        & unit
        & movable
        & $1$
        & path
        & {\em max}
        & 3D
        & NP-hard \cite{ppnph-os-99}
        \\ \hline
push
        & unit
        & movable
        & $1$
        & storage
        & {\em max}
        & 2D
        & NP-hard \cite{ppnph-os-99}
        \\ \hline \hline

{\bf push}
        & {\bf unit}
        & {\bf movable}
        & {\bf 1}
        & {\bf path}
        & {\bf {\em max}}
        & {\bf 2D}
        & {\bf NP-hard}
        \\ \hline \hline
push
        & unit
        & movable
        & $k$
        & path
        & min
        & 2D
        & open \cite{do-mpams-92}
        \\ \hline
push
        & unit
        & movable
        & $1$
        & path
        & min
        & 2D
        & open
        \\ \hline
\end{tabular}
\end{center}
\caption{Pushing block problems.}
\tablab{results}
\end{table}

\newpage
\small
\paragraph{Acknowledgements.}
We thank Therese Biedl for helpful discussions.
The third author acknowledges many insights from meetings of the
Smith Problem Solving Group.%
\normalsize
\footnote{
Beenish Chaudry,
Sorina Chircu,
Elizabeth Churchill,
Alexandra Fedorova,
Judy Franklin,
Biliana Kaneva,
Haley Miller,
Anton Okmianski,
Irena Pashchenko,
Ileana Streinu,
Geetika Tewari,
Dominique Thi{\'e}baut,
Elif Tosun.
}

\bibliographystyle{alpha}
\bibliography{/home1/orourke/bib/geom/geom}
\end{document}